\documentstyle[11pt,aaspp,epsf]{article}

\begin{document}

\title{Requirements for Investigating the Connection Between
Lyman Alpha Absorption
Clouds and the Large-Scale Distribution of Galaxies}
\author{Vicki L. Sarajedini}
\affil{Steward Observatory, University of Arizona, Tucson, AZ 85721}
\author{Richard F. Green}
\affil{NOAO\footnote{The National Optical Astronomy Observatories
are operated by the Association of Universities for Research in
Astronomy, Inc., under Cooperative Agreement with the National
Science Foundation.}, P.O. Box 26732, Tucson, AZ 85726}
\author{Buell T. Jannuzi\footnote{Hubble Fellow,
current address is NOAO, P.O. Box 26732, Tucson, AZ 85726}}
\affil{Institute for Advanced Study, Princeton, NJ 08540}

\begin{abstract}

We model the requirements on observational data that would allow an
accurate determination of the degree of association between
Lyman $\alpha$ absorbers and peaks in the redshift distribution
of galaxies (large-scale structures like clusters of galaxies).  We
compare simulated distributions of low-redshift Lyman $\alpha$
absorption systems, constrained to be consistent with the distribution
observed with HST, with the large-scale distribution of galaxies
determined from pencil-beam redshift surveys.  We estimate
the amount of observational data required from catalogues of
Lyman $\alpha$ absorbers and galaxies to allow a statistically
significant test of the association of absorbers with large-scale
structures of galaxies.

We find that for each line-of-sight observed for Ly$\alpha$ absorption
lines (assuming that the entire redshift range out to z$\simeq$0.4 is
observable),
redshifts must be obtained
for at least $\sim$18 galaxies
brighter than M$_B=-18$ and having redshifts between 0.2 and 0.4.
Based on the
redshift surveys used in this study, a search radius
of $\sim$10$\arcmin$ from the quasar line-of-sight is required.
This will ensure that all peaks in the galaxy redshift distribution are
represented by at least one galaxy in the observed sample.
If Lyman $\alpha$ absorbers are intrinsically uncorrelated
with galaxies, we find that $\sim$8 lines-of-sight must be observed to
show that the distributions are different at the 95\% confidence
level.  However, if a fraction of the Lyman $\alpha$ absorbers are
distributed with the peaks in the galaxy distribution, $\sim38$
lines-of-sight must be mapped for the distribution of both Lyman $\alpha$
absorbers and galaxies in order to determine the fraction of absorbers
distributed with the peaks of the galaxy distribution to an accuracy
of $10$\%.

\end{abstract}


\section{Introduction}

Ly$\alpha$ absorption clouds, observed in the spectra of quasars, are
numerous and detectable to high redshift.  As intervening absorption
systems, they are tracers of the evolution of the
gaseous content of the Universe.
By
understanding the extent to which these absorbers are associated with
clusters and large-scale structures of galaxies, we hope to trace the
evolution of these structures as well.

The connection between low redshift Ly$\alpha$ absorbers at z$\leq$0.5
and their high redshift counterparts is presently unclear.  The high
redshift Ly$\alpha$ absorbers display very little velocity
correlation.  They have been proposed to originate from intergalactic
clouds and, consistent with the absence of clustering, would not be
associated with galaxies \markcite{(Sargent et al. 1980)}.
Groundbased studies of these absorbers have shown strong evolution in
their number density \markcite{(e.g., Bechtold 1994)}.  The situation
at low redshifts may be different.  The study of Ly$\alpha$ absorbers
at low redshifts has been made possible through the use of the Hubble
Space Telescope (HST) to obtain ultraviolet spectra of moderate
redshift quasars.  These observations have revealed a larger number of
absorbers than was initially expected based on the extrapolation of
the evolution observed at higher redshifts \markcite{(Bahcall et
al. 1991; Morris et al. 1991; Bahcall et al. 1993a)}.
They appear to
be more clustered than high redshift absorbers, although to a lesser
degree than galaxies \markcite{(Bahcall et al. 1995)}.  The
relationship between Ly$\alpha$ clouds observed at low redshifts and
large-scale galaxy structures can in principle be determined directly
by comparing the observed distribution of the Ly$\alpha$ absorbers
detected in QSO spectra with the distribution of galaxies in the
fields of the same quasars.

There have been several examples of individual Ly$\alpha$ absorption
lines which appear to be associated with galaxies lying at the same
redshift.  Such matches have been found in the directions of the
quasars H1821+643 \markcite{(Bahcall et al. 1992)} and PKS 0405$-$123
\markcite{(Spinrad et al. 1993)}
with absorbers having EW $>$ 0.32 \AA.
The matches typically are found to
have impact parameters ranging from 70$h^{-1}$kpc to 160$h^{-1}$kpc
(h=H$_o$/100 km/s/Mpc)
and lie within 500 km/s of the Ly$\alpha$ absorption line redshift.
More recently, Lanzetta et al. (1995) have surveyed galaxy redshifts
in the fields of several quasars observed by HST and found 11
galaxy/absorption system matches with impact parameters
$\lesssim$160$h^{-1}$kpc.  They conclude that at least 0.32 $\pm$ 0.10
Ly$\alpha$ absorption lines are due directly to intervening galaxies
having large halo regions of hydrogen gas.
While such  studies will
help determine if some individual galaxies are associated with Ly
$\alpha$ absorbers, they do not allow us to investigate the relationship
between the absorbers and the large-scale structures of galaxies.  These
studies lack complete redshift sampling of the galaxies in the
individual quasar fields and cover limited regions (in
angular extent) around each quasar.  As a result, the complete
line-of-sight distribution of galaxies is not well determined on size
scales comparable to that of
clusters of galaxies and it
is not generally possible to determine if the observed galaxies are
part of larger structures (groups or clusters of galaxies).

There is evidence that Ly$\alpha$ clouds could be distributed
randomly with respect to regions of high
galaxy density.  In the direction of the quasar 3C273, one of the
strongest Ly$\alpha$ lines is found in a region where the nearest
luminous galaxy is 10 Mpc away \markcite{(Morris et al. 1993)}.  The
Morris et al. study also found that in a region of high galaxy
concentration in the same field, no Ly$\alpha$ lines are found within
36 Mpc.  Stocke et al. (1995) have made
similar findings using the CfA
galaxy redshift survey in the direction of Mrk 501.  They find no
galaxies brighter than M$_{B}$=-16 within 100$h{_{75}}{^{-1}}$ kpc
of a $>$ 4$\sigma$ Ly$\alpha$ absorption line detection.
Although the galaxy coverage is very good for these studies, they
probe a relatively small path length in redshift space and therefore
intercept fewer strong Ly$\alpha$ lines (observed equivalent widths
greater than 0.32\AA) that can then be compared with the galaxy
distribution. In general, all the Ly$\alpha$ lines studied in the
Stocke et al. work are weaker than the typical Ly$\alpha$ lines
observed and studied at higher redshift.

In this paper we investigate what data are necessary to adequately
determine the extent and nature of the association between the
large-scale galaxy structures and Ly$\alpha$ absorbers observed along
the sightlines to distant quasars. Obviously, catalogues of both
absorbers and the large-scale structures are needed. Not as clear is
the exact nature and size that these databases must take.  In the case
of the absorption line catalogue we will assume that for the near
future the largest and most complete catalogue available will be that
being compiled by the HST Quasar Absorption Line Key Project (Bahcall
et al. 1993a; Bahcall et al. 1995).  Their observations will provide a
large and homogeneous catalogue of strong Ly$\alpha$ absorption
lines toward over 80 quasars with redshifts between 0.15 and 1.9.
However, the number of Ly$\alpha$ absorbers along each individual
line-of-sight is greatly reduced from what we see at high redshift and
the entire redshift path length for each line-of-sight is not observed
for every quasar (in fact most of the quasars were not observed below
1600 \AA). Given this catalogue of Ly$\alpha$ absorbers, we have
conducted various simulations designed to determine what is required
of a redshift survey in the fields of these quasars in order to test
the following hypotheses: 1) low redshift Ly$\alpha$ absorbers are
uncorrelated with the large-scale structures of galaxies traced by the
peaks in the galaxy distribution; 2) some fraction of the Ly$\alpha$
absorbers is associated with the peaks in the galaxy distribution with
the remaining absorbers being unassociated.  Specifically, we
determine the characteristics and number of the galaxies that need to
have measured redshifts along each sightline and the total
number of QSO sightlines that must be studied in order either to
falsify the first hypothesis or to determine the fraction of associated
absorbers if the second hypothesis is valid.

\section{Analysis}

Several factors make the investigation of the relationship between
strong absorbers and large-scale
structures difficult at redshifts less than
0.2. First, there are extremely few strong Ly$\alpha$ absorption
systems at very low redshift (intrinsically, the volume density is low
as a practical consequence of the limited amount of path length
observed at redshifts less than 0.2). Second, the volume of space
surveyed at low redshift is relatively small, yielding few large-scale
structures for comparison. Third, those structures which are present
are difficult to identify without a very wide field (on the order of a
one degree diameter) redshift survey.  As a result, most of the
information in studying the relationship between absorbers and large
scale structures (at least for the strong lines contained in the Key
Project database) will come by comparing the cluster and group
distribution to the absorbers at redshifts between 0.2 and 0.5.

In this section we try to determine the rough characteristics (e.g. number
of galaxies in each field, how bright, over what redshift range, for
how many quasar fields) an incomplete redshift survey would need in
order to test either of the hypotheses stated at the end of the
introduction.

\subsection{Sampling the Galaxy Distribution}

If only a small number of galaxy redshifts are obtained
near the sightline of a quasar, the redshift distribution of
these galaxies does not show obvious peaks and voids and therefore does
not yield much information about the location in redshift space of
large-scale structures of galaxies.
We are interested in determining
the minimum number of galaxies which would allow us to best estimate
the location of large-scale structures in redshift space.
In this section, we determine the minimum number of
galaxies which must be observed near the quasar line-of-sight
satisfying the following
criteria:  1) the subset has the largest possible fraction of its
galaxies in the
``peak'' regions in redshift space and 2) every peak in the true galaxy
distribution is represented by at least one galaxy in the subset.

To model the effects of incomplete sampling of the true distribution
of galaxies we simulated limited observations of ``true''
distributions as defined from two very extensive galaxy redshift
surveys along different sightlines.  The components of these surveys
are described in Tables 1 and 2 \markcite{(Peterson et al. 1986;
Broadhurst et al. 1993; Colless et al. 1990; Broadhurst 1994)}.  One
survey is in the direction 1043+00 and contains a total of 213 galaxy
redshifts.  The pencil beam diameter of the survey is about
$\sim$32$\arcmin$ corresponding to 6.8$h^{-1}$Mpc at z=0.25.  The
survey samples galaxies down to an absolute magnitude of
M$_{B}\simeq$-18 out to z$\simeq$0.54 where we have
assumed H$_o$ = 75 km/s/Mpc.
The second survey is in the
direction of the South Galactic Pole at 0055-28 and contains 161
galaxy redshifts.  The diameter of the survey beam is about
$\sim$21$\arcmin$ corresponding to 4.6$h^{-1}$Mpc at z=0.25.  This
survey samples galaxies down to an absolute magnitude of
M$_{B}\simeq$-18 out to z$\simeq$0.46.  Both surveys include a much
broader cone covering about 3$\deg$, but extending to redshifts
of only $\sim$ 0.1.  The nominal completeness of the various components of each
survey is listed in the tables.  These surveys provide us with an
empirical test bed for modeling the limitations of incomplete redshift
surveys in representing the large-scale structures contained in the
survey volume.

Figures 1a and b show the redshift distributions of these
two surveys collapsed along a single line-of-sight in redshift space.
There are obvious peaks in the distributions noted by Broadhurst et
al. (1990) as the pencil-beam intersects large-scale structures.
The survey pencil-beam diameters are close to optimal for
detecting wall-like topologies on scales comparable to those revealed
in the CfA surveys \markcite{(Szalay et al. 1991)}.
Since we
are interested in the large-scale distribution of galaxies,
we have conducted our tests using the galaxy distribution in redshift
space alone without considering the effects of different impact
parameters of the individual galaxies to the QSO sightline.
Over the
angular fields we are considering, the distribution of the galaxies in
redshift space provides ample information about the locations of
groups and clusters of galaxies along a sightline.  Our tests were
designed to simulate actual ``observations'' of galaxies along these
lines-of-sight through randomly selected subsamples of the total data
set representing the objects whose redshifts are obtained in
a given limited redshift survey.
Using these surveys as a representation of the ``true'' universe, we
will investigate how well selected subsets of the sample represent
the large-scale distribution of the galaxies including the incidence
of clusters or peaks in the redshift distribution.

Our first step is to define a weighting function in redshift space
that indicates association with peaks in the redshift surveys.
For the redshift distributions of Figures 1a and b,
we have chosen a histogram representation with
bin size of $\Delta$z = 0.01 so that a typical galaxy cluster
or group in the data would be sampled by at least 2 bins.
The location and width of the peaks were
determined mainly through visual inspection
based on apparent over-densities in the galaxy distribution
and the width of those over-densities at half of their maximum height
in the histogram.
The peak locations and widths can also be identified by determining
the local noise level (standard deviation)
within a 3 bin radius of the assumed peak.
Gaussian statistics were used to determine the standard deviation from 6
bins consisting of 3 bins on either side of the peak, excluding those
associated with nearby peaks.
Our ``peaks'' are those bins which are
$\ge$3$\sigma$ above this noise level.
Once the location and width of peaks in each galaxy distribution
has been determined, we can define our weighting function.
The weighting function is designed so that regions in redshift space that
are ``associated'' with a peak have a weight of 1.0 and regions that are
well outside the peaks have weights of 0.  We can consider each peak
as a Gaussian shape having a full-width at half-maximum equal to the
peak width.  The ``peak'' regions in our
weighting function are then defined as the peak center $\pm$ 1$\sigma$
and the function value within these regions is 1.0.  Beyond $\pm$
1$\sigma$ the function value behaves as a Gaussian, trailing off towards
0 in the ``void'' regions of redshift space (see Figures 2a and b).  In
this way, each peak has a finite width in the weighting function.

The weighting function can be used as a measure of the degree of
concentration of any subsample of galaxies to the peaks.
To do so, the galaxies in a subsample are assigned an initial delta function
of unit amplitude, then weighted by multiplication with the galaxy
distribution functions in Figures 2a and b for each
of the two redshift surveys respectively.
The mean function value for the weighted sample measures the concentration
in peaks of the redshift distribution, or the averaged probability that
an individual galaxy in the sample has a redshift associated with a peak.
Obviously, not all galaxies in
the redshift surveys fall within a peak according to the weighting function
of Figure 2.
To satisfy the first criterion mentioned at the beginning
of this section, we investigated what constraints could be placed
on subsample selection to allow a higher fraction of the
galaxies to fall within the ``peaks''.
These constraints will point to a sampling strategy for determining
the redshift peaks in newly observed samples with the best attainable
reliability.

Figures 1a and b show the galaxy distributions with
all of the galaxies from each of the smaller surveys providing us with
our ``true'' map of the universe.  The low z peaks are more
easily identified
due to the inclusion of the 3$\deg$ diameter
field which
extends to only z$\lesssim$0.1.  However, the deep portion of these
surveys, necessary in order to consider comparison
between the absorbers and large-scale structures, only covers the
inner 20$\arcmin$ to 30$\arcmin$ of each survey.  The hatched region
in Figures 1a and b represents the galaxy distribution within this
smaller cone.  When only the smaller angular field is considered, it
becomes difficult to identify the low redshift peak in the
distribution.
Since the data set we are using to represent
the ``true'' universe does not allow us to define a peak at
low redshift, and for the reasons described at the beginning of this
section, we place a lower limit of z=0.2 for
identifying peaks in cones of 30$\arcmin$ or less.
An upper limit at z=0.4 is imposed by the
limitations for obtaining redshifts for a reasonable sampling of the
galaxy luminosity function with 4-m class
telescopes.  Therefore, we have little information from the two
surveys used here about the true galaxy distribution above z=0.4 and
cannot determine how well galaxies beyond this redshift represent
the peaks and voids of the distribution.

Intrinsically bright galaxies have a somewhat higher probability
of being in the
peaks of the distribution.
Supporting evidence is found in the fact that eliminating the data
from the redshift surveys having z $<$ 0.2
increases the mean function value determined from the remaining
galaxies for both redshift surveys.
The less luminous objects
observable at low redshifts, seem to be more uniformly distributed.
Limiting the observed
galaxy absolute magnitude range for a random sampling of the galaxies
in the field improves the probability that a given galaxy is a member
of a peak in the true galaxy distribution.  The redshift surveys we
are using have galaxy absolute magnitudes ranging from M$_B\simeq$ -18
to M$_B\simeq$ -21 based on their apparent magnitudes with H$_o$=75
km/s/Mpc.  If we exclude the few galaxies which are fainter than M$_B$
= -18, the concentration in peaks is increased.  While raising
this lower limit may increase the concentration further it also greatly
decreases the number of observable galaxies, i.e.  there aren't enough
luminous galaxies available to locate and define the peaks in the
galaxy distribution.  We therefore have chosen to constrain the range
of galaxy absolute magnitude to M$_B$ $\leq$ -18 which is $\sim$
0.44$L^\star$ \markcite{(Marzke et al. 1994)}. This corresponds to an
apparent magnitude of $B$ $\leq$ 23 at z=0.4.  Eliminating galaxies
with M$_B$ $>$ -18 increases the mean function value from the
remaining galaxies in each of the two redshift surveys.

Once we have
maximized the fraction of galaxies falling within the peaks
in an ``observed'' subset,
we can determine the minimum size subsample that retains
this same fraction of galaxies associated with peaks in
the galaxy distribution.
In addition, these subsets must also have
at least one galaxy in the redshift range of each peak in the
true galaxy distribution.  In this way, we can make certain that
all peaks in the true galaxy distribution are represented.

To do this,
we simulated ``observations'' of the true universe by randomly selecting
galaxies in the test surveys
through sub-cones 5$\arcmin$ in diameter using numerous cones to span
each survey out to the largest effective angular size of the
survey.  For our purposes, simulating observations of galaxies at
significant redshift (out to z=0.4), the ``effective angular size'' is
constrained to be the largest angle in each test survey for which
galaxy redshifts out to at least z=0.4 are available.  For this reason
we limit our simulations to the the inner 35$\arcmin$ x 32$\arcmin$
of the 1043+00
survey and the inner 23$\arcmin$ x 10$\arcmin$ of the
South Galactic Pole survey.
This same simulated observational procedure was repeated with
increasing subcone sizes until
the maximum size of the survey was reached.

For each ``observation'', the galaxies found within the cone were
assigned an intial delta function of unit amplitude, then weighted by
multiplication with the galaxy distribution function in Figures 2a and b.
By averaging the weighted amplitudes of
all of the galaxies observed within a cone, a mean function value is
determined.  For example, if half of the galaxies in a cone fall
within 1$\sigma$ of the central redshift of a peak and the other half
fall completely outside, the mean function value for that cone is 0.5,
indicating a 50\% chance that a given galaxy in that subset was
selected from a peak in the true galaxy distribution.

Figures 3a and b show the number of galaxies observed in a cone vs.
the mean function value for each of the two redshift surveys with the
data sampling limits as described above.  Each dot represents a
different sub-cone of the total survey within which $n$ galaxies have
been observed.  These plots contain cone sizes from 5$\arcmin$ in
diameter to the largest angular size possible within the survey
limits; 35$\arcmin$ x 32$\arcmin$ for the 1043+00 survey and
23$\arcmin$ by 10$\arcmin$ for the SGP survey.  It is clear that if
all of the galaxies in these surveys which are between the stated
redshift and magnitude limits are observed, then the mean function
value is 0.68 for the 1043+00 survey and 0.74 for the South Galactic
Pole survey.  As we examine galaxy sets containing fewer and fewer
observed galaxies, the typical deviation from the mean function value
for a given subset becomes
greater than $\sim$5\% for subsets containing fewer than
$\sim$18 galaxies.  For samples with sizes below this limit,
the mean function value, expressing the probability that a given galaxy in
that subset lies within 1$\sigma$ of a peak in the true distribution,
becomes very uncertain.

These simulations of galaxy observations along a specific
line-of-sight suggest that the probability of observed galaxies lying
in redshift peaks reaches a maximum between 0.68 and 0.74 where we
have limited our redshift range to 0.2$\leq$z$\leq$0.4 and
placed a lower limit on the absolute magnitude of M$_B\leq$ -18
corresponding to an apparent magnitude of $B\leq$ 23 at z=0.4.
A subsample of
at least $\sim$18 galaxy redshifts are needed for a representative
sample where
$\sim$70\% of the galaxies fall within 1$\sigma$ of a peak in
the true galaxy distribution.  The angular field of view necessary to
obtain this minimum number of galaxy redshifts is r $\simeq$
11$\arcmin$ in the 1043+00 survey and r $\simeq$ 7$\arcmin$ in the
South Galactic Pole survey.  These spatial ranges correspond to
sampling regions of space $\sim$2.2 Mpc in size at
z=0.3.

The angular diameters quoted here, however, are dependent on the
completeness and efficiency for the redshift surveys we used.
The selection efficiency for obtaining galaxy redshifts is critical
in determining the angular diameter of the cone required to
measure a sufficient number of objects.  A simple integration of the
galaxy luminosity function
\markcite{(Marzke et al. 1994)}, in a truncated cone of r = 10$\arcmin$
and redshift limits of 0.2 to 0.4 suggests that we would find $\sim$120
galaxies mith M$_B\leq$ -18, about six times the number of galaxies
within the same constraints for the surveys used here.
It is therefore possible to limit the
angular area required for search around each quasar by increasing the
efficiency with which redshifts are obtained.  There is a lower limit
set by the physical area subtended by the large-scale structures
themselves.  One might consider a Mpc or so as the minimum diameter
below which the search becomes more relevant to individual objects
rather than clusters or associations.  A field with a radius of
4$\arcmin$ at z=0.2 would encompass over 1 Mpc
and provide
enough galaxies for redshift measurement to meet our criterion
given a $\sim$100\% efficiency for obtaining redshifts.  With
a $\sim$60\% efficiency, the minimum number of galaxy redshifts could be
obtained within a 5$\arcmin$ radius field.

There are $\sim$10-12 sets containing 18 or more galaxies in each
of the two redshift surveys.  These subsamples
can then be binned in redshift space in the same manner as the entire
redshift survey.  If we consider each galaxy in these subsets as a
peak, we find that each peak, defined from the total galaxy
distribution within the 0.2$\leq$z$\leq$0.4 redshift range, is
represented.  By assuming that each galaxy represents a peak in the
galaxy distribution, all true peaks are located.  However, only
$\sim$70\% of the galaxies are associated with true peaks;
approximately 30\% of the galaxies will actually lie in the ``void''
regions of the true galaxy distribution.  We find that overestimating
the number of peaks by 30\% is
the minimum error which can be attained in defining
peaks in redshift space while also minimizing the total
number of galaxy redshifts obtained.

\subsection{Comparing the Galaxy and Lyman Alpha Cloud Distributions}

Comparison of the distributions of galaxies and absorbers requires not
only the sample of galaxies, for which the determination of
large-scale structure was discussed in the previous section, but also
a sample of absorbers.  For this paper we will assume that low
redshift Ly$\alpha$ absorption systems detected on lines-of-sight to
quasars at redshifts greater than 0.4 are consistent in distribution
and number with what has been observed by the HST Quasar Absorption
Line Survey (Bahcall et al. 1993a).  Therefore, determining the number
of Ly$\alpha$ clouds necessary to test the two hypotheses stated in
the introduction can be restated as determining the minimum number of
lines-of-sight that need to be observed for absorbers and galaxies.

We are assuming that the absorption lines found along
a line-of-sight will be drawn from the simplified line distribution
function
\begin{equation}
{{dN}\over{dz}}={({{dN}\over{dz}})}_o{(1+z)}^\gamma
\end{equation}
with (dN/dz)$_o = 18$ and $\gamma = 0.3$ as determined from HST
observations by Bahcall et al. (1993a) based on the detection
of Ly$\alpha$ lines having rest equivalent widths of 0.32\AA
or greater.
To model the case of no association between large-scale structure
and absorbers, simulated samples of Ly$\alpha$
absorbers were generated with no velocity correlations
on small scales and with a line-of-sight
density evolution with redshift as defined above.
Eq. 1 was used to determine the probability that a line would exist
($\Delta$N) within a certain redshift bin ($\Delta$z).  The redshift
bin size was chosen to be the
resolution element size for the
HST Quasar Absorption Line Survey which is $\Delta$v=270 km/s.
We then divided the redshift space between 0.2 and 0.4 into bins
of this size.
A random number generator was used to produce a number between 0 and
1 for each bin.  If that number was less than the probability
$\Delta$N determined for that bin, a Ly$\alpha$ line would be generated at
that redshift.
This same procedure was repeated to simulate line lists from many
lines-of-sight with the total distribution with redshift consistent with
the global distribution found by Bahcall et al.

To obtain a quantitative measure of association between the peaks
in the galaxy distribution and absorbers,
the redshift distribution of each random absorption
line list was then weighted by the
galaxy redshift distribution function (see Figures 2a and b)\footnote
{All tests comparing the Ly$\alpha$ line distributions to that of
galaxies along a sightline were performed separately using both redshift
surveys.  Because the results from each survey were consistent with
one another, we will only present results
for the 1043+00 survey in the text and figures.}.
We compare the distribution of Ly$\alpha$ line
function values to similarly computed distributions of function values
for subsets of galaxy redshifts drawn from the 1043+00 and SGP
surveys.
The number of galaxies along each sightline was chosen in a manner
consistent with the discussion in section 2.1, providing the minimum
number of galaxy redshifts to represent the peaks in the galaxy
distribution.  The dotted line in Figure 1 shows the arbitrarily
normalized galaxy selection function for the 0.2$\leq$z$\leq$0.4
range for each survey based on a Schechter luminosity function
(Marzke at al 1994).  The effects of this function are ignored in
our simulations when choosing galaxy subsets for comparison with
the Ly$\alpha$ line lists
since variations of this function are small over this redshift range.

As the number of sightlines
increases, the number of galaxies in the comparison sample
must also increase.  Since the previous experiment indicates that
$\sim$18 galaxies are necessary to characterize adequately the galaxy
distribution along a single sight line,
we conducted this test assuming that the minimum number
of galaxy redshifts will be measured for each sightline.
A set of 18 galaxies is compared with the Ly$\alpha$ line list for a
single sight line,
a set containing 36 galaxies is compared with 2 Ly$\alpha$ line lists, etc.
Since the galaxies in these distributions are drawn from the redshift
range of z=0.2 to 0.4, the Ly$\alpha$ line lists generated contain
lines within this range which results in an average of 3.9 absorbers
per line list.

The Kolmogorov-Smirnov test (KS test) was used to find the level of
confidence at which the distribution of weighting function values for observed
galaxies is different from that of the locally Poissonian distribution
of Ly$\alpha$ lines.  Figure 4 shows this KS
probability as the number of observed sightlines increases.  To show
that the two distributions are different at the 90\% confidence level,
at least 6 lines-of-sight must be observed or $\sim$24 Ly$\alpha$
lines in total.
At least 8 lines-of-sight are
necessary for 95\% confidence and 12 are required to show that they
are different at the 99\% confidence level.
This exercise demonstrates that the distribution of
Ly$\alpha$ lines in redshift space, constructed to
show locally Poissonian statistics,
obviously differs from a representative distribution of galaxies which
shows clustering.  If the Ly$\alpha$ lines are distributed in this
way, we would need to study only $\sim$8 to 12 sightlines for
absorbers (observed at the level of the HST Key Project observations)
and galaxies (with a limited redshift survey as discussed above) in
order to recognize that the absorbers and large-scale structures of
galaxies are not distributed in the same manner.

Next we explored the possibility that only a fraction of Ly$\alpha$
lines are randomly distributed while the remainder are associated with
the peaks in the galaxy distribution. We
compared the degree of concentration in the peaks for
Ly$\alpha$ line lists generated from the power-law equation
to that for Ly$\alpha$
lines distributed like the peaks in the galaxy distribution.
Remember, however, that unless we have a very large number of galaxy
redshifts along the quasar sightline, we cannot know the location of
actual peaks in the galaxy distribution with absolute certainty.  Our
earlier simulations indicate that with subsets of at least $\sim$18 galaxies
(meeting the selection criteria detailed above and assuming each
of these galaxies is located in a ``peak'') we overestimate the number
of peaks by 30\%.
The weighting function values (which measure the degree of association with
peaks) for these generated Ly$\alpha$ lines
must therefore be corrected for a 30\% excess in apparent
associations.  We chose to do that in the simulations by generating an
``associated'' line list from the ``observed'' galaxy sample, then
applying the ``true'' weighting function to assign zero weight
to the 30\% which are spurious associations.

The peak associated Ly$\alpha$ lines are generated in much the same way
as those with global properties characterized by Eq. 1 as described earlier
in this section.  The only difference is that we have substituted the
probability function of Eq. 1 with
an empirically determined probability function representing
the peaks.  In other words, rather than using the pure power-law
equation for dN/dz in Eq. 1 to determine the probability that a line
will be placed in a particular $\Delta$z redshift bin, we are using the
global trend for dN/dz modified by the location and width of
peaks as determined
by subsets of galaxies containing at least $\sim$18 galaxies.
This distribution assumes the same number density of low-redshift
Ly$\alpha$ lines as described by Eq. 1 but
redistributed on small scales to correlate with the peaks.

The Monte-Carlo generated Ly$\alpha$ lines were then weighted by the
true galaxy distribution functions in Figures 2a and b so that each line
received a value based on its redshift location with respect to true
peaks in the galaxy distribution.  Figure 5 shows the normalized
cumulative distributions of these function values for 10000 Ly$\alpha$
lines distributed with no association to the peaks in the galaxy
distribution (solid line) and 10000 Ly$\alpha$ lines distributed like
the peaks in the galaxy distribution (dashed line).
Within the distribution of random Ly$\alpha$ lines there are many that
fall at or
near a function value of 0 since many random lines will fall
between peaks based on our weigting function in Figures
2a and b.  For the peak-distributed lines (dashed line), note that
fewer fall at 0 and many more have function values of 1.0 since they
have been generated to be associated with peaks in the galaxy distribution.
Still, some have function values at or near 0 due to the fact that the
peak-associated lines have been generated based on peak locations as
defined from subsets of $\sim$18 galaxies.  Since only 70\% of these
galaxies actually fall in a true peak, we have generated some
peak-associated lines at redshifts where an actual peak doesn't
exist.  These lines will have values at or near 0 since they don't fall
in the ``true'' peak regions in redshift space.
Any mixture of the two distributions in Figure 5 will
produce a cumulative distribution located between them determined by
the percentages of each parent population contained within it.

The aim of our test is to characterize to what accuracy we can
determine the composition of an observed distribution of Ly$\alpha$ lines
taken from several lines-of-sight based on the proximity of each
Ly$\alpha$ line to peaks in the galaxy distribution.  In other words,
if we obtain a sample of Ly$\alpha$ line redshifts, how accurately can
we determine the fractional contribution of each of the two parent
distributions, purely unclustered lines and galaxy peak associated
lines, and what is the minimum number of Ly$\alpha$ lines which must
be observed to make this determination?
Our approach differs from that taken by
Bahcall et al (1993a) who calculated the number of Ly$\alpha$ lines
necessary to detect clustering within the Ly$\alpha$ sample alone.
We are interested in determining the number of Ly$\alpha$ lines needed
to identify a population clustered with galaxies
when we are able to make use of additional information about the
distribution of the galaxies along the same sight lines as
the absorbers.
For our simulation the
distribution of galaxies
is determined from the simulated limited
redshift surveys described in section 2.1.

Our approach is to generate samples of lines for which the fractional
contribution of each parent distribution is known.  The sample is then
compared through a KS test to a set of distributions with a range in
fractional composition of the two parents
(Figure 5).  This computation produces a distribution of KS probability
values peaking at the fractional mixture of the two parents which best
fits the sample.  Since the KS test determines the probability that
two distributions are different, a large KS probability indicates a
lack of difference or a ``best fit''.

Many random draws of samples containing the same number of lines, for
example, in a 50:50 ratio will produce a range of KS probability
distributions where the best fit will vary around $\sim$50\%.  To take
into account the variation in peak location and width of the KS
probability distribution, we generated distributions for 100 random
draws for each sample of Ly$\alpha$ lines and determined the mean
fractional composition value and its variance at the KS probability
peak.  The range in fractional composition of parent populations which
includes 95\% of the random draws is represented by the mean peak
value $\pm$ 2 times the standard deviation of the distribution.  We
computed this range for samples containing various total numbers of
Ly$\alpha$ lines and fractional compositions.  We find that the
standard deviation for Ly$\alpha$ line sets containing the same total
number of lines remained roughly the same regardless of composition
value.

Figure 6 reveals that as more Ly$\alpha$ lines are considered, the
range of fractional composition values decreases allowing for the true
composition to be more accurately determined.  At 150 Ly$\alpha$
lines, approximately 38 lines-of-sight, enough Ly$\alpha$ absorption
lines are observed within the 0.2$\leq$z$\leq$0.4 redshift
range to determine the fractional composition to within 10\% of the
true value for $\sim$95\% of the generated samples.  The accuracy
improves slowly as the number of lines-of-sight increases.  If fewer
than 5 sightlines are observed, these tests suggest that it is
impossible to determine the composition of the sample, i.e what fraction
are associated with large-scale structures if some fraction are random
with respect to these structures.  5 sightlines are equivalent
to only $\sim$20 Ly$\alpha$ lines in the redshift range being considered.
We find that
the measurement of Ly$\alpha$ absorbers from at least 38 sightlines is
necessary (equivalent to 150 Ly$\alpha$ lines), in addition to the
minimum number of galaxy redshifts
needed, to show that the population of Ly$\alpha$ absorbers is
composed of a mixture of those which are distributed like the peaks in
the galaxy distribution and those which are uncorrelated with the
peaks in the galaxy distribution.

\section{Discussion}

Recent surveys to identify and obtain redshifts
for galaxies projected near quasar
sightlines have been designed to study the relationship between absorbers and
individual galaxies.  Other surveys have concentrated on the association of
quasars with host galaxy clusters.  Often the field of view is limited by the
cassegrain spectrograph with multi-object capability.  Such samples can serve
as the starting point for defining the larger-scale structures.  They may not
be adequate to do so in their current form; they were not designed to be.

Two examples from current surveys illustrate the point.
The quasar PKS 0405-123 has been observed with HST revealing 14 Ly$\alpha$
absorbers within the redshift range of z=0.081 to 0.540
\markcite{(Bahcall et al. 1993b)}.  Recently, Ellingson et al. (1994)
published a list of 29 galaxy redshifts in this field ranging from
z=0.16 to 0.66.  This redshift survey covers a field 5.9$\arcmin$ by
3.8$\arcmin$ around the quasar and is 78\% complete to r=21.5.  The
redshift range of this survey is comparable to the surveys used in our
significance tests.  In the optimal redshift range determined for the
surveys used in our simulations (z=0.2 to 0.4), 10 galaxy redshifts
are measured which meet our absolute magnitude criterion of
M$_B\leq$-18. We assume B-R colors for these galaxies of up to 1.75
(Colless et al. 1990).  Figure 3 indicates that a sample of 10
galaxy redshifts does not meet the minimum sample size requirements to
ensure that 70\% of the galaxies will
fall within 1$\sigma$ of a peak in the
galaxy redshift distribution.

As another example, consider the available observations of the field
of the quasar 3C~351.
This quasar has also been observed with HST revealing $\sim$16
Ly$\alpha$ lines within the redshift range of z=0.092 to 0.370
\markcite{(Bahcall et al. 1993a)}.  Included in the survey
of Lanzetta et al. (1995) are redshifts for 10 galaxies in this field,
ranging from z=0.07 to 0.370.  Their survey of this field covers
$\sim$ 5$\arcmin$ in diameter and is 57\% complete down to r=21.5.
Within their sample, 4 galaxies meet our redshift range and absolute
magnitude criteria.  Again, the small sample size does not ensure that
70\% of these galaxies
lie within 1$\sigma$ of a large-scale
structure.

More extensive redshift coverage of galaxies in wider fields around
these quasars would allow for a better determination of the
large-scale distribution of galaxies along the sightline.  Although
some observations of galaxies in the fields of many more of the HST
observed quasars exist, not enough redshifts have been measured to
yield statistically significant results.  Ideally, it would be
necessary to obtain enough galaxy redshifts along the line-of-sight to
a quasar having a redshift at or beyond z=0.4 to satisfy the
requirements determined in section 2.1.  In this way, all the peaks in the
galaxy distribution out to the QSO redshift would be represented with
an overestimation of 30\%.
We would then need to obtain these surveys in
the fields of at least $\sim$8 QSO's for which HST spectra are
available in order to test the hypothesis that the Ly$\alpha$
absorption clouds are uncorrelated with the peaks in the galaxy
distribution.  It would be necessary to obtain these surveys in the
fields of $\sim$38 QSO's with HST spectra to determine
to 10\% accuracty what
fraction of the Ly$\alpha$ absorbers are associated with the peaks in
the galaxy distribution.

\section{Conclusions}

We have conducted various numerical experiments designed to
investigate the significance with which the association between
Ly$\alpha$ absorption clouds and the large-scale distribution of
galaxies can be determined.  We have found that in pencil-beam
redshift surveys extending to redshifts of z$\sim$0.5, the maximum
probability of selecting a subsample of galaxies such that every peak
in the distribution is represented by at least one galaxy occurs when
redshifts for at least 18 galaxies are obtained between
z=0.2 and 0.4 with M$_B\leq$=-18 and drawn from an angular radius of
$\sim$10$\arcmin$ around the quasar line-of-sight.
The limited cone size
of the surveys used in these simulations is the main factor pushing us
to the z$\geq$0.2 region to sample volumes of space large enough to
detect large-scale structures.  Based upon these redshift surveys, it
would be necessary to have at least $\sim$18 galaxy redshifts in each
quasar field to populate all the peaks in the galaxy distribution
along the line-of-sight.  Without knowing the true galaxy distribution
along a sightline, we conclude that $\sim$70\% of the $\sim$18 galaxy
redshifts measured will fall within $\sim$1$\sigma$ of a true peak in the
galaxy distribution.
A typical sightline must be surveyed in a cone of radius
$\sim$10$\arcmin$ to get this number of redshifts down to $B$=23 at
z=0.4 for the surveys used in this study, typical of a 4-meter class
telescope redshift survey.

If the Ly$\alpha$ absorption clouds are uncorrelated with the peaks in
the galaxy distribution, we find that at least 8 lines-of-sight must
be observed to show that the distribution of galaxies and that of the
absorbers is different at the 95\% significance level.  However, if
some fraction of the Ly$\alpha$ absorbers is distributed like the
peaks in the galaxy distribution and some fraction is uncorrelated, we
find that $\sim$38 lines-of-sight must be observed to determine the
fraction (to 10\% accuracy) of absorbers which are distributed like
the galaxies.

Our test results clearly indicate that more data are needed in order
to draw reliable conclusions about the extent and nature of the
association between Ly$\alpha$ absorption clouds and the peaks in the
galaxy distribution along the line-of-sight.  Fortunately, several
research groups are actively obtaining redshifts for galaxies in the
fields of quasars observed with HST.

\acknowledgments
We would like to thank Tom Broadhurst for providing the redshift
surveys used in this paper in electronic form.  Thanks also to Gary
Schmidt, Joe Shields, Jill Bechtold and Ata Sarajedini for helpful
conversations.  We thank Simon Morris for his valuable comments
and suggestions for improving this paper.  B.T.J. acknowledges
support for this work by NASA
through grant number HF-1045.02-93A from the Space Telescope Science
Institute, which is operated by the Association of Universities for
Research in Astronomy, Incorporated, under NASA contract NAS5-2655.

\clearpage

\begin{planotable}{cccccccc}
\tablewidth{33pc}
\tablecaption{Redshift Survey Data for 1043+00 Field}
\tablehead{
\colhead{\# of Galaxies}      &
\colhead{z range}          & \colhead{$B$ range}  &
\colhead{field size}          & \colhead{completeness} &
\colhead{ref.}}
\startdata
70& 0.0 to 0.097& 13.95 to 17.19& 209$\arcmin$.3 x 213$\arcmin$.2& 74\%& 1 \nl
53& 0.0 to 0.213& 17.32 to 19.69& 34$\arcmin$.3 x 33$\arcmin$.5& $\sim$80\%& 4
\nl
108& 0.0 to 0.438& 19.70 to 20.77& 36$\arcmin$.8 x 31$\arcmin$.8&
$\sim$80\%& 4 \nl
32& 0.0 to 0.543& 21.05 to 22.50& 5$\arcmin$.3 x 12$\arcmin$.0& 81\%& 3 \nl
\end{planotable}

\begin{planotable}{lrrrrcrrrrr}
\tablewidth{33pc}
\tablecaption{Redshift Survey Data for South Galactic Pole Field}
\tablehead{
\colhead{\# of Galaxies}      &
\colhead{z range}          & \colhead{$B$ range}  &
\colhead{field size}          & \colhead{completeness} &
\colhead{ref.}}
\startdata
75& 0.0 to 0.133& 13.82 to 17.50& 208$\arcmin$.8 x 209$\arcmin$.0& 52\%& 1 \nl
59& 0.0 to 0.444& 20.50 to 21.50& 22$\arcmin$.8 x 9$\arcmin$.5& 84\%& 2 \nl
27& 0.0 to 0.564& 20.71 to 22.42& 5$\arcmin$.3 x 12$\arcmin$.0& 80\%& 3 \nl
\tablerefs{
(1) Peterson et al. 1986; (2) Broadhurst, Ellis \& Shanks 1993;
(3) Colless et al. 1990;
(4) Broadhurst 1994.}
\end{planotable}

\clearpage

\centerline{FIGURE CAPTIONS}

Figure 1 - The redshift distribution of galaxies in the 1043+00 survey
(a) and the South Galactic Pole survey (b).
The hatched region
represents the galaxies within the inner 35$\arcmin$ by 32$\arcmin$
for the 1043+00 survey and the inner 23$\arcmin$ by 10$\arcmin$ for
the SGP survey.  These are the regions used in the simulations.
The dotted line is the arbitrarily normalized selection function
for galaxies based on a Schechter luminosity function (see text in
Section 2.2).

Figure 2 - Peak-defining function based on the distributions for each
of the two surveys in Figure 1.  The location and width of each peak
was determined by smoothing the galaxy distribution to find the
average background galaxy level and defining the peaks as any points
which fell above 1$\sigma$ of this background.  Regions of redshift
space within 1$\sigma$ of a peak in the galaxy distributions of Figure
1 have function values of 1.0.  Beyond 1$\sigma$, the function value
behaves as a Gaussian.

Figure 3 - Number of observed galaxies vs. the mean function value for
each of the two surveys in Figures 1 and 2.  The mean function value
is an average of the $n$ observed galaxies' function values as
determined from Figure 2.

Figure 4 - KS probability that the distribution of function values for
simulated Ly$\alpha$ absorption lines is different from the
distribution of function values for a subset of galaxies as a function
of the number of sightlines observed.  For a single sightline,
$\sim$3.9 Ly$\alpha$ lines are observed in the redshift range 0.2 to
0.4.  A small KS probability indicates that the two distributions are
different.

Figure 5 - Cumulative distributions of the function values for
Monte-Carlo generated Ly$\alpha$ lines distributed as a power-law with
no small scale velocity correlation (solid line) and Ly$\alpha$ lines
distributed like the peaks in the galaxy distribution as defined from
the minimum number of galaxies necessary in an incomplete redshift
survey ($\sim$18) (dashed line).  The function value is determined for
each simulated Ly$\alpha$ line from the function shown in Figure 2.

Figure 6 - Twice the standard deviation in the fractional composition
of a sample of Ly$\alpha$ lines as a function of the number of
Ly$\alpha$ lines in the sample (see text).  At $\sim$150 Ly$\alpha$
lines (corresponding to $\sim$38 lines-of-sight) enough absorption
lines are observed within the z=0.2 to 0.4 redshift range to determine
the fractional composition to within $\pm$ 10\% of the true value for
$\sim$95\% of the generated samples.

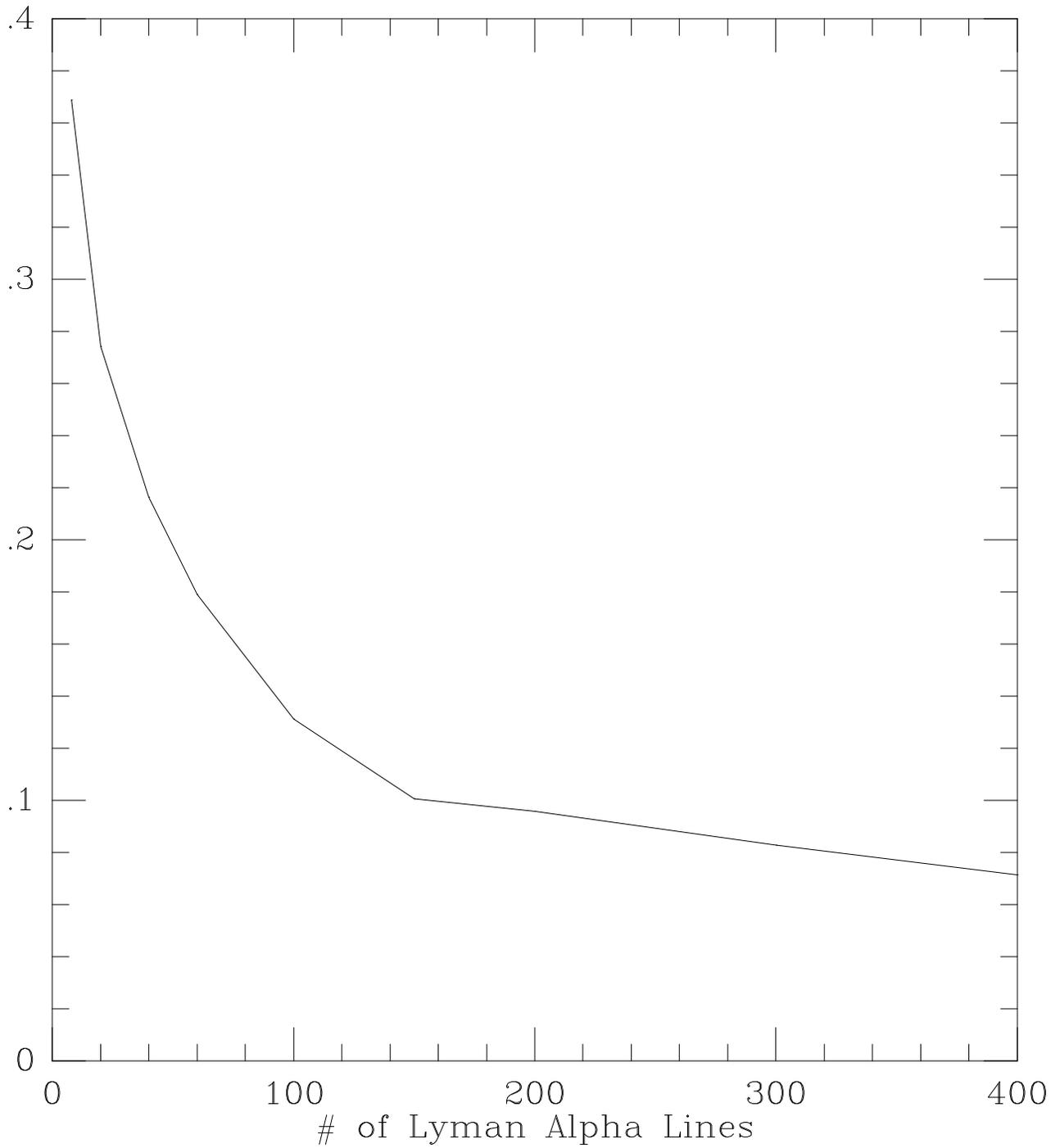
\end{document}